\documentclass[referee]{raa}           

\usepackage{graphicx,times}
\usepackage{natbib}
\usepackage{amssymb,amsmath}
\usepackage{float}
\bibpunct{(}{)}{;}{a}{}{,}

\usepackage[pagebackref=true]{hyperref}

\begin{document}

   \title{Dark matter search in dwarf irregular galaxies with ten years of data from the IceCube neutrino observatory}

 \volnopage{ {\bf 20XX} Vol.\ {\bf X} No. {\bf XX}, 000--000}
   \setcounter{page}{1}

   \author{Yi-Fei L{\"u} \inst{1}, Ben-Yang Zhu \inst{1}, Rong-Lan Li \inst{1}, Xue-Kang Guo \inst{1}, Tian-Ci Liu \inst{1}, Yong-Bo Huang \inst{1}, Yun-Feng Liang  \inst{1}
   }


   \institute{Guangxi Key Laboratory for Relativistic Astrophysics, School of Physical Science and Technology, Guangxi University, Nanning 530004, China; {\it liangyf@gxu.edu.cn}\\
\vs \no
   {\small Received 20XX Month Day; accepted 20XX Month Day}
}

\abstract{Dwarf irregular galaxies (dIrrs), as rotationally-supported systems, have more reliable J-factor measurements than dwarf spheroidal galaxies and have received attention as targets for dark matter detection in recent years. 
In this paper, we use 10 years of IceCube muon-track data and an unbinned maximum-likelihood-ratio method to search for neutrino signals beyond the background from the directions of 7 dIrrs, aiming to detect neutrinos produced by heavy annihilation dark matter. We do not detect any significant signal. Based on such null results, we calculate the upper limits on the velocity-averaged annihilation cross section for 1 TeV - 10 PeV dark matter. Our limits, although weaker than the strictest constraints in the literature in this mass range, are also a good complement to the existing results considering the more reliable J-factor measurements of dIrrs.
}

   \authorrunning{Y.-F. L{\"u} et al. }            
   \titlerunning{Search for DM in dIrrs with IceCube}  
   \maketitle

%
\section{Introduction}           
\label{sect:intro}
Astrophysical and cosmological evidence has suggested the existence of particles beyond the Standard Model, dark matter (DM), which account for about 84\% of the matter component of the universe (\citealt{2018RvMP...90d5002B,2016A&A...594A..13P,2020A&A...641A...6P,2021PrPNP.11903865A}). Although its nature is still unknown, many models, such as primordial black holes (PBH) 
 (\citealt{PBH2021arXiv211002821C}), weakly interacting massive particles (WIMPs) (\citealt{2018RPPh...81f6201R,2005PhR...405..279B}), axion/axion-like particles (ALPs) (\citealt{2017ehep.confE..83S}) and sterile neutrinos (\citealt{2021arXiv210900767K}) are proposed as dark matter candidates. Recent years, most dark matter searches focus on the mass range of the sub-eV (e.g., ALPs) or the GeV-TeV scale (e.g., WIMPs) and have not found a significant signal (\citealt{2023ApJ...945..101A}). The situation prompts us to consider the models in a wider mass range, one example is the ultra-heavy dark matter (UHDM).

The UHDM has a mass of above 10 TeV, which means it is too massive to be produced in current colliders. There are already many viable candidates for UHDM (\citealt{2022arXiv220306508C,1998PhRvD..59b3501C,2016PhLB..760..106B,2016PhRvD..94i5019B,2017PhRvD..96j3540K}), and in this paper we focus on the annihilation UHDM which has received less attention before. In the scenario of thermal production of dark matter, the cross section of DM annihilation to Standard Model particles scales as $M^{-2}_{\chi}$ (\citealt{2022ApJ...938L...4T}). As the mass increases, the cross section generally decreases. If it becomes too small, then the DM will not be sufficiently depleted before freeze-out, leading to an inconsistency with the observed cosmological DM density. This unitarity consideration limits the DM mass to be below $\thicksim$ 100 TeV (\citealt{2017JHEP...02..119B,1990PhRvL..64..615G}). Although this unitarity boundary has long hindered people's interest and enthusiasm in searching for the annihilation UHDM, several mechanisms have been proposed that can violate this boundary (\citealt{2022arXiv220306508C,2019PhRvL.123s1801K}). For instance, if the DM exists in a composite state rather than a point-like particle, thermal DM with PeV-scale masses could be possible (\citealt{2022ApJ...938L...4T,2016JHEP...08..151H}). In such a senario, the unitarity bound can be violated because the DM annihilation cross section no longer scales as $M^{-2}_{\chi}$. Instead, it is now related to the geometric size of the composite DM.

One way to detect dark matter particles is to observe the gamma-rays, neutrinos from the DM annihilation or decay in the astrophysical environments. Promising sites for the observation of neutrinos from dark matter annihilation include the Galaxy clusters, Galactic halo and center, as well as the cores of the Sun and Earth (\citealt{2016arXiv160105691Z}). Among other targets,  dwarf spheroidal galaxies (dSphs), one type of the Milky Way satellite galaxies, are also considered as promising objects because of their relatively close proximity and clean astrophysical background (\citealt{2015ICRC...34.1226W}). However, the stellar kinematics of these pressure-supported galaxies are highly uncertain, which makes it difficult to determine their DM profiles very well (\citealt{2013NewAR..57...52B}). The accurate determination of the DM profile is very important for estimating astrophysical factor (J-factor) of DM indirect detection (\citealt{2016MNRAS.461.2914H}). 

Dwarf irregular galaxies (dIrrs) are targets that we have paid less attention to (\citealt{2021PhRvD.104h3026G,2023ApJ...945...25A,2019ICRC...36..520H,2021PhRvD.103j2002A}), and they have some properties similar to dSphs, such as high DM abundance and low astrophysical background. Unlike dSphs, dIrrs are located further away and they are rotationally supported star-forming dwarf galaxies (\citealt{2021PhRvD.104h3026G,2015AJ....149..180O}) (in contrast,dSphs are pressure-supported systems). Their DM profiles can be obtained from their rotation curves. Meanwhile, dIrrs are in general larger galaxies compared to dSphs and have larger dark matter halos (their halo masses are $M_{200}= 10^{7}-10^{10}M_{\odot}$), and may host more DM subhalos (\citealt{2017MNRAS.465.4703K}). The standard $\Lambda$CDM cosmological model predicts that dark matter halos are filled with many self-bound substructures, which has been confirmed by N-body simulations (\citealt{2016A&A...594A..13P,2012AnP...524..507F}). The presence of subhalos is expected to significantly enhance the DM signal (\citealt{2022JCAP...06..023R,2022JCAP...10..021L}). Therefore, the enhancement factor due to subhalos increases the detectability of the dIrrs. These relevant properties make dIrrs also interesting targets for DM searches. 

\begin{table}[t]
\caption[]{Sample of dIrr galaxies. Columns: (1) galaxy name (2) the right ascension (3) declination (4) galaxy distance (5) the angle to virial radii (6) source extension (7) the astrophysical factor for annihilation (8) halo mass (9) Boost Factor (10) TS value.
\label{tab1}}
\centering
\renewcommand\arraystretch{1}
\setlength{\tabcolsep}{1.2mm}{
\centering
\begin{tabular}{lccccccccc}
\hline
\hline
\rm{Name} & $\begin{array}{c}\text{\rm{Ra}} \\ (\mathrm{deg})\end{array}$ & $\begin{array}{c}\text{\rm{Dec}} \\ (\mathrm{deg})\end{array}$ & $\begin{array}{c}\text{\rm{D}} \\ (\mathrm{Mpc})\end{array}$ & $\begin{array}{c}\theta_{vir} \\ (\mathrm{deg})\end{array}$ & $\begin{array}{c}\sigma_s \\ (\mathrm{deg})\end{array}$ & $\begin{array}{c}\log_{10}J \\ (\mathrm{GeV}^{2}\mathrm{~cm}^{-5})\end{array}$ & $\begin{array}{c}\log_{10}M_{200} \\ (M_{\odot})\end{array}$ & $\begin{array}{c}\ \rm{Boost Factor}\end{array}$ & $\begin{array}{c}\ \rm{TS}\end{array}$ \\
\hline
\textit{\rm{IC10}} & 5.10 & 59.29 & 0.79 & 3.857 & 0.078 & 17.857 & 10.6 & 5.53 & 0.04 \\
\textit{\rm{IC1613}} & 16.19 & 2.13 & 0.76 & 2.361 & 0.120 & 17.632 & 10.0 & 5.01 & 0.03 \\
\textit{\rm{NGC6822}} & 296.23 & -14.80 & 0.52 & 5.325 & 0.159 & 18.173 & 10.5 & 5.45 & 0.00 \\
\textit{\rm{WLM}} & 0.49 & -15.46 & 0.98 & 2.609 & 0.082 & 18.062 & 9.6 & 4.72 & 0.00 \\
\textit{\rm{DDO133}} & 188.22 & 31.54 & 4.88 & 0.784 & 0.033 & 17.501 & 10.5 & 5.45 & 0.13 \\
\textit{\rm{DDO154}} & 193.52 & 27.15 & 4.04 & 0.695 & 0.030 & 17.800 & 10.3 & 5.25 & 0.00 \\
\textit{\rm{DDO168}} & 198.61 & 45.91 & 4.25 &1.142 & 0.032 & 17.365 & 10.9 & 5.86 & 0.46 \\
\hline
\hline
\end{tabular}}
\end{table}

Previous works by IceCube collaboration, which have searched for dark matter signals from the Galactic center (\citealt{2020PhRvD.102h2002A,2021JInst..16C9009I}), Galactic halo (\citealt{2023PhRvD.108j2004A,2023arXiv230210542D}), Sun center (\citealt{2022PhRvD.105f2004A,2019arXiv190903930A}), Earth center (\citealt{2019ICRC...36..541R,2022PhRvD.105h3025S}), and nearby galaxies and galaxy clusters (\citealt{2013PhRvD..88l2001A,2022icrc.confE.506J}), found no significant excess beyond background expectations. In our previous work, we have used IceCube observations on dwarf spheroidal galaxies to search for dark matter (\citealt{2023PhRvD.108d3001G}). In this paper, we further search for DM in dwarf irregular galaxies on the basis of the previous work. Different from the previous analysis, we analyze dwarf irregular galaxies as extended sources and consider the enhancement of the annihilation signal due to their internal subhalos. We will search for UHDM induced neutrinos from dwarf irregular galaxies (dIrrs) using 10 years (April 2008 to July 2018) of Icecube muon track data (\citealt{2021arXiv210109836I}) and an unbinned maximum-likelihood-ratio method (\citealt{2008APh....29..299B}). We select seven dIrrs, four sources (WLM, NGC6822, IC10, IC1613) are the most promising objects for DM detection as proposed in previous works (\citealt{2018PhRvD..98h3008G,2021PhRvD.104h3026G}), and the remaining three sources are selected from (\citealt{2023ApJ...945...25A}). Note that the IceCube 10-years muon track data have a much higher sensitivity to the northern hemisphere sky. Nevertheless, we have also included two sources in the southern sky (WLM and NGC6822) in the sample to demonstrate this sensitivity difference. The sample information is shown in Table~\ref{tab1}.


\section{SIGNAL EXPECTATIONS}
\label{sect:2}

The expected differential neutrino flux for the annihilation of dark matter particles is calculated as follows:
\begin{equation}
\frac{d\Phi_{\nu}}{dE_{\nu}}\left(E_{\nu}\right)=\frac{\langle\sigma v\rangle}{8 \pi m_{\chi}^{2}} \frac{dN_{\nu}}{dE_{\nu}}\left(E_{\nu}\right) J_{\left(\mathrm{\Delta\Omega}\right)}\times B
\label{eq:eq1}
\end{equation}
where $\langle\sigma v\rangle$ is the velocity-averaged annihilation cross section, $m_{\chi}$ is the mass of the dark matter particle, ${dN_{\nu}}/{dE_{\nu}}$ is the annihilation spectrum of DM particles, and $B$ is the boost factor.

\begin{figure}[t]
    \centering
    \includegraphics[width=0.6\textwidth,height=0.45\textwidth]{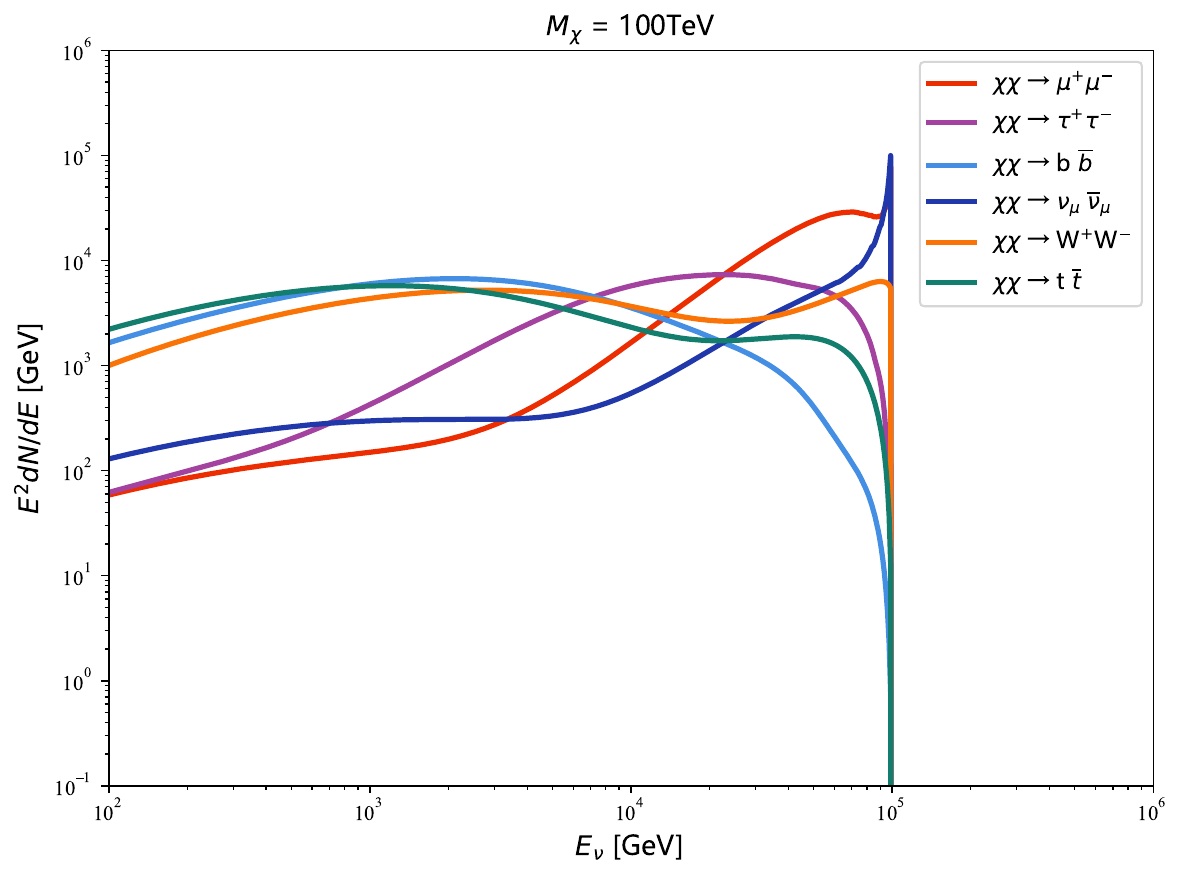}
    \caption{Spectra of neutrinos from the annihilation of DM particles with a mass of 100 TeV for 6 channels, assuming that the branching ratio for each channel is 100\%.}
    \label{Figure:dnde}
\end{figure}

We assume dark matter annihilates into the standard model particles with a 100\% branch ratio and calculate the energy spectrum of neutrinos ${dN_{\nu}}/{dE_{\nu}}$. For this calculation we use the HDMSpectra (\citealt{2021JHEP...06..121B}), a Python package providing spectra of various annihilation and decay channels of heavy dark matter. In this work, we consider the DM masses ranging from 1 TeV to 10 PeV and 6 annihilation channels: $b\bar{b}$ , $t\bar{t}$ , $\tau^{+}\tau^{-}$, $\nu_{\mu}\bar{\nu}_{\mu}$ , $\mu^{+}\mu^{-}$ and $W^{+}W^{-}$. Fig.~\ref{Figure:dnde} shows the neutrino spectra for the annihilation of a 100 TeV DM through different channels.

The flux is proportional to the J-factor (the integral over the square of the dark matter density),
\begin{equation}
J_{\left(\mathrm{\Delta\Omega}\right)}=\int_{\mathrm{\Delta\Omega}}\int_{\mathrm{l.o.s}}dld\Omega\rho_{\rm{DM}}^{2}(r(l))
\label{eq:eq2}
\end{equation}
where $l$ is the distance along the line of sight (l.o.s.) and $r(l)=\sqrt{D^{2}-2lD\cos\theta+l^{2}}$ with $D$ the distance to the source and $\theta$ the angular separation between the directions of the source center and l.o.s.  In our work, we adopt the J-factors reported in (\citealt{2023ApJ...945...25A}).

The universal rotation curve model (\citealt{1996MNRAS.283.1102P}) can fit the dIrr rotation curve well (\citealt{2018PhRvD..98h3008G}), and the DM distribution in these galaxies is well described by a Burkert profile (\citealt{1995ApJ...447L..25B}):
\begin{equation}
\rho_{\rm{DM}}(r)=\rho_{0}\frac{r^3_{0}}{\left(r_{0}+r\right)\left(r^2_{0}+r^2\right)}
\label{eq:eq3}
\end{equation}
where $\rho_{0}$ is the normalization density, and $r_{0}$ is the core radius.

The substructures within dIrrs can increase the expected DM flux. We parameterize this effect with the boost factor $B$ in Eq.(\ref{eq:eq1}). For the calculation of the boost factors we mainly refer to the formulas and results in (\citealt{2017MNRAS.466.4974M}). They considered the VL-II (\citealt{2008Natur.454..735D}) and ELVIS (\citealt{2017MNRAS.464.3108G,2014MNRAS.438.2578G}) N-body cosmological simulations to study the properties of subhalos, obtaining the concentrations of the subhalos as a function of the distance $R_{\rm{sub}}$ to the main halo center. Together with the concentration-mass relation derived in (\citealt{2011PhRvD..83b3518P}) for the field halo, we can calculate the luminosity $\mathcal{L}_{\rm{smooth}}(\rm{M})$ of the smooth halo for a given halo mass ${M}$ and the luminosity $\mathcal{L}_{\rm{smooth}}(m,x_{\rm{sub}})$ of the subhalo for a given halo mass $m$ at the distance $R_{\rm{sub}}$ ($x_{\rm{sub}}=R_{\rm{sub}}/R_{\rm{vir}}$ with $R_{\rm{vir}}$ the virial radius). Then the boost factor can be calculated with: 
\begin{equation}
B(M)=\frac{3}{\mathcal{L}_{\rm{smooth}}(M)}\int_{M_{\rm{min}}}^{M}\frac{dN(m)}{dm}dm\int_{0}^{1}dx_{\rm{sub}}\left[1+B(m)\right]\mathcal{L}(m,x_{\rm{sub}})x^2_{\rm{sub}}
\label{eq:eq4}
\end{equation}
The $dN(m)/dm$ is the subhalo mass function, which reads $dN(m)/dm=A/M(m/M)^{-\alpha}$. The normalization factor is equal to $A = 0.012$ for a slope of the subhalo mass function of $\alpha= 2$ and to $A = 0.03$ for $\alpha= 1.9$ (\citealt{2014MNRAS.442.2271S}). It is important to note that there are large uncertainties in the calculation of the boost factor $B(\rm{M})$, which depends on the index $\alpha$ of the mass function, the assumed minimum subhalo mass $M_{\rm{min}}$, the concentration-mass relation adopted for subhalos and whether or not the tidal stripping effect being taken into account. In this work, we adopt the median value of the boost factors reported in (\citealt{2017MNRAS.466.4974M}): the mass function index is assumed to be 1.9, the minimum halo mass is $10^{-6}M_{\odot}$, and the tidal stripping effect is not considered. The boost factors are shown in Table~\ref{tab1}.

\section{ICECUBE DATA ANALYSIS}

IceCube is a cubic-kilometer neutrino detector located at the South Pole. It consists of 5160 optical photosenors, which are connected to the data acquisition system via cables (“strings”). Neutrinos interact with nuclei and electrons in the ice producing charged particles and emitting Cherenkov light, which is detected by the photosensors (\citealt{2017JInst..12P3012A}). The geometry and sensitivity of the photosensors lead to an effective low-energy threshold of about 100 GeV for neutrinos (\citealt{2003ICRC....3.1369Y}) .

In this work, we utilize the IceCube updated public data of muon tracks. This data set consists of muon tracks observed by IceCube from April 2008 to July 2018, including a total of 1134450 muon-track events. We use all IC40 to IC86-VII data, with the number in the name corresponding to the number of installed detector strings. The data includes direction and energy information of neutrino events, as well as instrument response functions (IRFs, including effective area, smearing matrix, etc) for different periods over the ten years (\citealt{2021arXiv210109836I}).
We process the $19\times2$ double-counted tracks in the dataset found in (\citealt{2022PhRvD.105i3005Z}), following the procedure in (\citealt{2022arXiv221003088C}) \footnote{See \url {https://github.com/beizhouphys/IceCube_data_2008--2018_double_counting_corrected} for the updated dataset.}.

The analysis is performed by using an unbinned maximum likelihood ratio method (\citealt{2008APh....29..299B,2011ApJ...732...18A,2022PhRvD.106h3024L,2013ApJ...779..132A,2017ApJ...835...45A,2022JCAP...12..002P}). The likelihood function is defined as follows:
\begin{equation}
\mathcal{L}\left(n_{s}\right)=\prod_{k}\prod_{\emph{i}=1}^{N^{k}}\left[\frac{n_{s}^{k}}{N^{k}} S_{\emph{i}}^{k}+\left(1-\frac{n_{s}^{k}}{N^{k}}\right) B_{\emph{i}}^{k}\right],
\label{eq:eq5}
\end{equation}
where $S_{\emph{i}}$ and $B_{\emph{i}}$ are the signal and background probability density functions (PDFs), respectively, $N^{k}$ is the total number of events in the sample $\emph{k}$ and $n^{k}_{s}$ is the number of signal events which is dependent on the model parameters. The superscript $\emph{k}$ refers to the $k$th IceCube data sample (IC40-IC86VII).

The signal and background PDFs for the $i$th event are functions of the reconstructed direction $x_{\emph{i}}$ and the reconstructed muon energy $E_{\emph{i}}$. For time-integrated searches the signal PDF $S_{\emph{i}}$ is given by:
\begin{equation}
S_{\emph{i}}=S_\emph{i}^{\rm{spat}}\left({x}_{\emph{i}}, \sigma_{\emph{i}} \mid {x}_{s}\right) \times S_\emph{i}^{\rm{ener}}\left(E_{\emph{i}} \mid \gamma\right),
\label{eq:eq6}
\end{equation}
where $S_\emph{i}^{\rm{spat}}$ is the spatial contribution which depends on the angular uncertainty of the event $\sigma_{\emph{i}}$ and the angular separation between the directions of the event ${x}_{\emph{i}}$ and the source ${x}_{s}$.
We model the spatial term of the signal PDF as a two-dimensional Gaussian,
\begin{equation}
S_\emph{i}^{\rm{\rm{spat}}}\left({x}_{\emph{i}}, \sigma_{\emph{i}} \mid {x}_{s}\right)=\frac{1}{2 \pi \sigma_{\emph{i}}^{2}} \exp \left(- \frac {{\mid x_{\emph{i}}-x_{s} \mid}^2}{2 \sigma_{\emph{i}}^{2}}\right)
\label{eq:eq7}
\end{equation}
where $x_{\emph{i}}$ is the direction of the $i$th event, and $x_{\emph{s}}$ is the direction to the source. Given the larger size of the dIrrs (compared to dSphs), it is better to analyze them as extended sources to get more realistic limits. In Table~\ref{tab1}, we present the extension corresponding to the virial radius and the 68\% containment angle of these dIrrs. When searching for an extended source, the value of $\sigma_{\emph{i}}$ is replaced with $\sigma = \sqrt{\sigma_{\emph{i}}^{2}+\sigma_{s}^{2}}$ (\citealt{2014ApJ...796..109A,2020ApJ...892...92A}), where $\sigma_{s}$ quantifies the extension of the source. In this work, we adopt the angle containing 68\% of the total emission $\left(J_{\left(\sigma_{s}\right)}= 0.68 \times J_{tot}\right)$ as $\sigma_{s}$ (\citealt{2022PhRvD.106l3032D,2015ApJ...801...74G}). The signal PDF becomes,
\begin{equation}
S_{\emph{i}}^{\rm{spat}}=\frac{1}{2 \pi \left(\sigma_{\emph{i}}^{2}+\sigma_{s}^{2}\right)} \exp \left(- \frac {{\mid x_{\emph{i}}-x_{s} \mid}^2}{2 \left(\sigma_{\emph{i}}^{2}+\sigma_{s}^{2}\right)}\right).
\label{eq:eq8}
\end{equation}

The energy term of the signal PDF, $S_\emph{i}^{\rm{ener}}$, is a function of the reconstructed muon energy proxy $E_{\emph{i}}$. Assuming a power law spectrum of spectral index $\gamma$ and for a given declination $\delta_{\emph{s}}$ of the source \footnote{Considering the small extension of the dIrrs we are studying (see $\sigma_{s}$ in Table~\ref{tab1}), we disregard the variations of the IRFs within the region covered by the source, and adopt the instrument response toward the center of the source ($\alpha_{s},\delta_{s}$).}, the PDF is (\citealt{2022MNRAS.514..852H}),
\begin{equation}
S_{\emph{i}}^{\rm{ener}}=\frac{\int E^{-\gamma}_{\nu} A_{\mathrm{eff}}^{k}\left(E_{\nu}, \delta_{s}\right) M_{k}\left(E_{\emph{i}} \mid E_{\nu}, \delta_{s}\right) d E_{\nu}}{\int E^{-\gamma}_{\nu} A_{\mathrm{eff}}^{k}\left(E_{\nu}, \delta_{s}\right) d E_{\nu}}
\label{eq:eq9}
\end{equation}
For the search for DM signals, the $E^{-\gamma}_{\nu}$ should be replaced by a DM spectrum (i.e., Eq.~(\ref{eq:eq1})).

The background PDF $B_{\emph{i}}$ is obtained directly from the experimental data and is given by
\begin{equation}
B^{k}_{\emph{i}}=B^{\rm{spat}}_{\emph{i}}\left(\delta_{\emph{i}}\right) \times B^{\rm{ener}}_{\emph{i}}\left(E_{\emph{i}} \mid \delta_{\emph{i}}\right)
\label{eq:eq10}
\end{equation}
The spatial term, $B^{\rm{spat}}_{\emph{i}}$, is the normalized event number per unit solid angle and is varied with the declination $\delta$ (\citealt{2021PhRvD.103l3018Z}),
\begin{equation}
B_{\emph{i}}^{\rm{spat}}\left(\delta_{\emph{i}}\right)=\frac{N_{\delta_{\emph{i}} \pm 3}^{k}}{N_{k} \times \Delta \Omega_{\delta_{\emph{i}} \pm 3}}
\label{eq:eq11}
\end{equation}
where $N_{\delta_{\emph{i}} \pm 3}$ is the number of events in the declination band of $\delta_{\emph{i}} \pm 3$ and $\Delta \Omega_{\delta_{\emph{i}} \pm 3}$ is its corresponding solid angle, $N_{k}$ is the total number of events in the sample $k$. The background PDF is right ascension independent because the IceCube observatory is located at the South pole.

The $B_{\emph{i}}^{\rm{ener}}$ represents the energy term of the background PDF, 
\begin{equation}
B_{\emph{i}}^{\rm{ener}}\left(E_{\emph{i}} \mid \delta_{\emph{i}}\right)=\frac{N_{\emph{i}\emph{j}}^{k}}{N_{\delta_{\emph{i}} \pm 3}^{k} \Delta E}
\label{eq:eq12}
\end{equation}
where $N_{\emph{i}\emph{j}}^{k}$ is the number of events within the declination range of $\delta \in\left[\delta_{\emph{i}}-3^{\circ}, \delta_{\emph{i}}+3^{\circ}\right)$ and the muon energy range of $\log _{10} E_{\mathrm{rec}} \in\left[\log _{10} E_{\emph{j}}, \log _{10} E_{\emph{j}}+0.1\right)$ for the sample $k$.

We maximize the likelihood in Eq.~(\ref{eq:eq5}) to derive the best-fit model parameters. A likelihood ratio test statistic of the signal is defined as,
\begin{equation}
\mathrm{TS}= -2 \log \left[ \frac{L\left(\hat{n}_{s}\right)}{L\left(n_{s}=0\right)}\right]
\label{eq:eq13}
\end{equation}
where $n_{s}$ is the number of signal events, and $\hat{n}_{s}$ is the best-fit value of the parameter.

For searching for DM signals, the number of signal neutrinos  from DM annihilation in Eq.~(\ref{eq:eq5}) is related to the DM parameters through,
\begin{equation}
n_{s} = T_{\rm{live}}\int_{E_{\rm{min}}}^{E_{\rm{max}}} A_{\mathrm{eff}}\left(E_{\nu}, \delta\right) \frac{d\Phi_{\nu}}{dE_{\nu}}\left(E_{\nu}\right)dE_{\nu}
\label{eq:eq14}
\end{equation}
where $T_{\rm{live}}$ is the detector uptime, $A_{\mathrm{eff}}$ is the effective area of the detector, $d\Phi_{\nu} / dE_{\nu}\left(E_{\nu}\right)$ is the energy spectrum of dark matter annihilating into neutrinos given by Eq.~(\ref{eq:eq1}), and the $E_{\rm{min}}$ and $E_{\rm{max}}$ are set to the the minimum and maximum energies allowed by the effective area provided in the 10-year IceCube data.

\section{RESULT}

\begin{figure}[h]
    \centering
    \includegraphics[width=0.6\textwidth,height=0.45\textwidth]{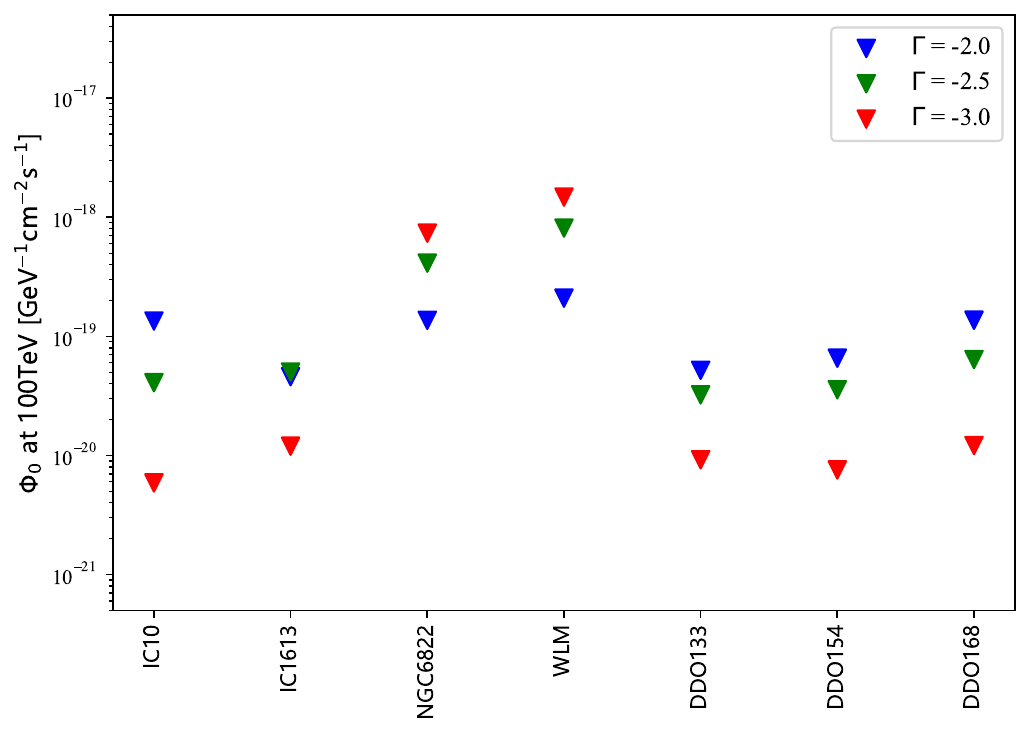}
    \caption{The 95\% C.L. upper limits on the neutrino flux from the 7 dIrrs at 100 TeV assuming a source spectrum of $E^{\Gamma}$}
    \label{Figure:flux}
\end{figure}

Assuming a power-law spectrum, the TS values of the putative neutrino signals from these dIrrs are listed in Table~\ref{tab1}. We do not find any significant excess beyond the background expectation in the directions of the 7 sources within our sample. So we derive the flux upper limits at the 95\% confidence level, assuming a source energy spectrum of $E^{-2}$, $E^{-2.5}$ and $E^{-3}$, respectively. The results are shown in Fig.~\ref{Figure:flux}.

\begin{figure}[t]
    \centering
    \includegraphics[width=15.0cm]{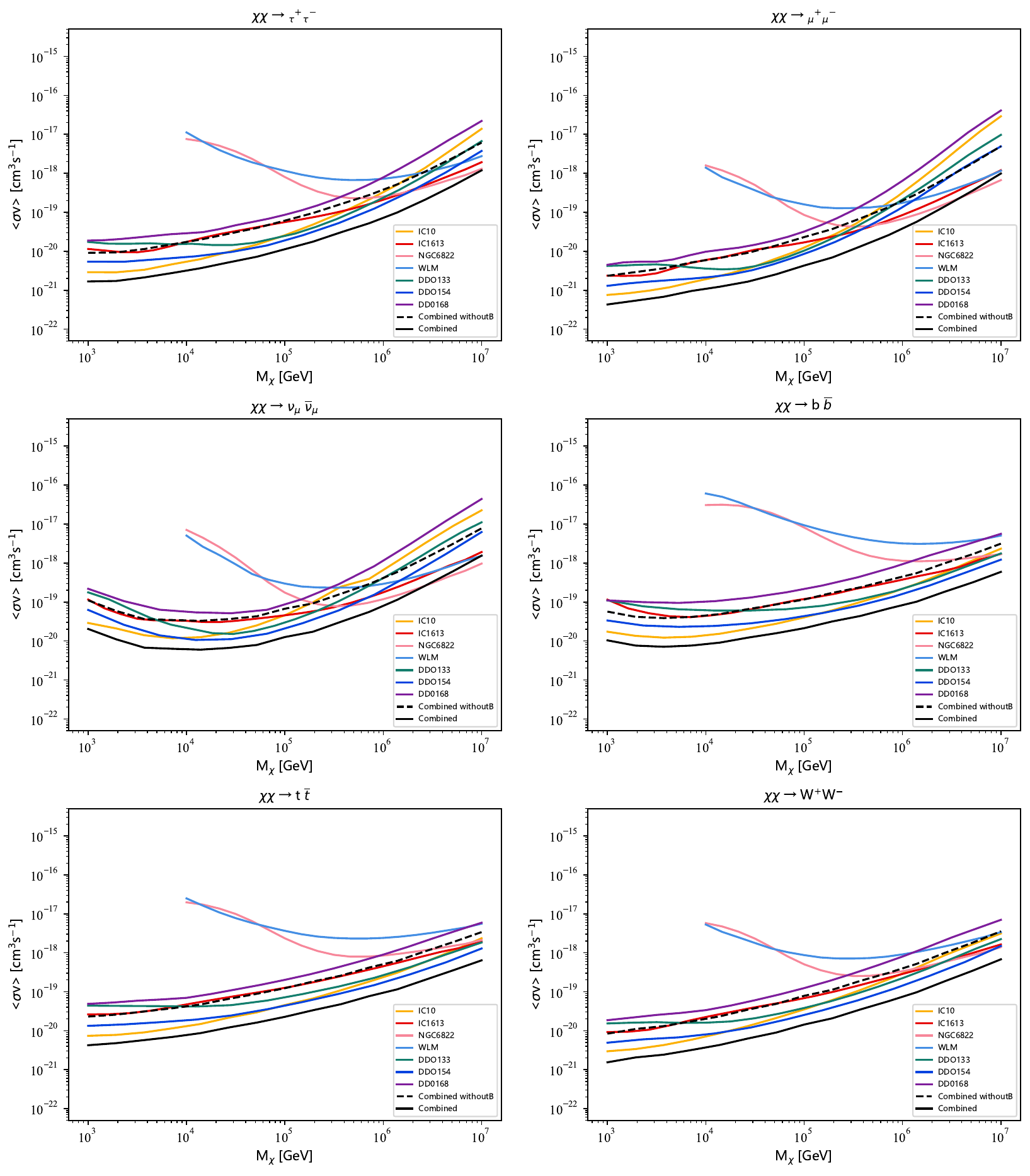}
    \caption{Upper limits on the DM annihilation cross section from the IceCube observations of dIrr galaxies for different channels. The black lines indicate the upper limits from the combined analysis of dIrrs. The two southern-hemisphere sources (WLM and NGC6822) give weaker constraints especially in the lower mass regime.}
    \label{Figure:result}
\end{figure}

Next, we explore the neutrino signals from the annihilation of dark matter. We use Eq.~(\ref{eq:eq1}) to perform searches for the DM mass range from 1 TeV to 10 PeV and different annihilation channels, and find no annihilation signal. We calculate the upper limits at the 95\% confidence level in the $\langle\sigma v\rangle-m_{\chi}$ parameter space for each dIrr. In Fig.~\ref{Figure:result} we show the limits for the 6 annihilation channels ($b\bar{b}$ , $t\bar{t}$ , $\tau^{+}\tau^{-}$, $\nu_{\mu}\bar{\nu}_{\mu}$, $\mu^{+}\mu^{-}$, $W^{+}W^{-}$). The strongest constraints we obtain are for the $\chi\chi \rightarrow \mu^{+}\mu^{-}$ channel.

 In Fig.~\ref{Figure:result}, for the two southern sky sources, it can be seen that they place overall much weaker limits on the $\left<\sigma v\right>$ than the dIrrs in the northern sky. However, at large DM masses ($\gtrsim10^6\,\rm GeV$), they give stronger constraints. Such results are reasonable. In the southern hemisphere, since the event rate of atmospheric muons is orders of magnitude higher than that of neutrinos \citep{2017JInst..12P3012A}, harsh cuts on the reconstruction quality and minimum energy are applied in order to suppress this atmospheric muon background, and the cuts remove most southern hemisphere events with an estimated energy below $\simeq10$ TeV \citep{2021arXiv210109836I} (so for NGC 6822 and WLM we only show limits for DM masses of $\textgreater$ 10 TeV in Fig.~\ref{Figure:result}). In contrast, for the northern hemisphere, we can filter out the atmospheric muon contamination by the Earth.
For neutrinos at the high-energy end ($>10^6\,\rm GeV$), on the other hand, events from the northern sky may interact far from the detector while crossing the Earth, resulting in unobservable energy losses and thus difficult to reconstruct. This leads to a significantly smaller effective area for the northern sky than the southern sky at $>10^6\,\rm GeV$ energies (one can see Fig.~2 of Ref.~\citealt{2021arXiv210109836I}). Since the energy spectra of the atmospheric muon and neutrino fluxes are expected to be softer than the astrophysical/DM neutrinos, their contribution to the background becomes less significant at these high energies. Therefore, for the large mass DM, the analysis of the two dIrrs (NGC6822, WLM) gives more strong constraints on the $\left<\sigma v\right>$.

To improve the sensitivity, we also jointly analyze the data from the 5 northern-sky sources simultaneously. In this case, the combined likelihood function is
\begin{equation}
\widetilde{\mathcal{L}}\left(\langle\sigma v\rangle,m_{\chi}\right) = \prod_{\emph{j}}\mathcal{L}_{\emph{j}}\left( n_{s}\left(\langle\sigma v\rangle,m_{\chi},J_{\emph{j}}\right)\right)
\label{eq:eq15}
\end{equation}
with $\mathcal{L}_{\emph{j}}\left(n_{s}\right)$ the likelihood in Eq.~(\ref{eq:eq5}) for the $j$th source. The details of the method of the combined analysis can be found in (\citealt{2023PhRvD.108d3001G,2023PhRvD.108h3034Z}). The limits from the combined analysis are shown as black lines in Fig.~\ref{Figure:result}. The results of the combined analysis are about 2.5 times stronger than the ones by DDO154, which gives the strongest constraints among the 7 dIrr galaxies.

\begin{figure}[h]
    \centering
    \includegraphics[width=0.7\textwidth,height=0.5\textwidth]{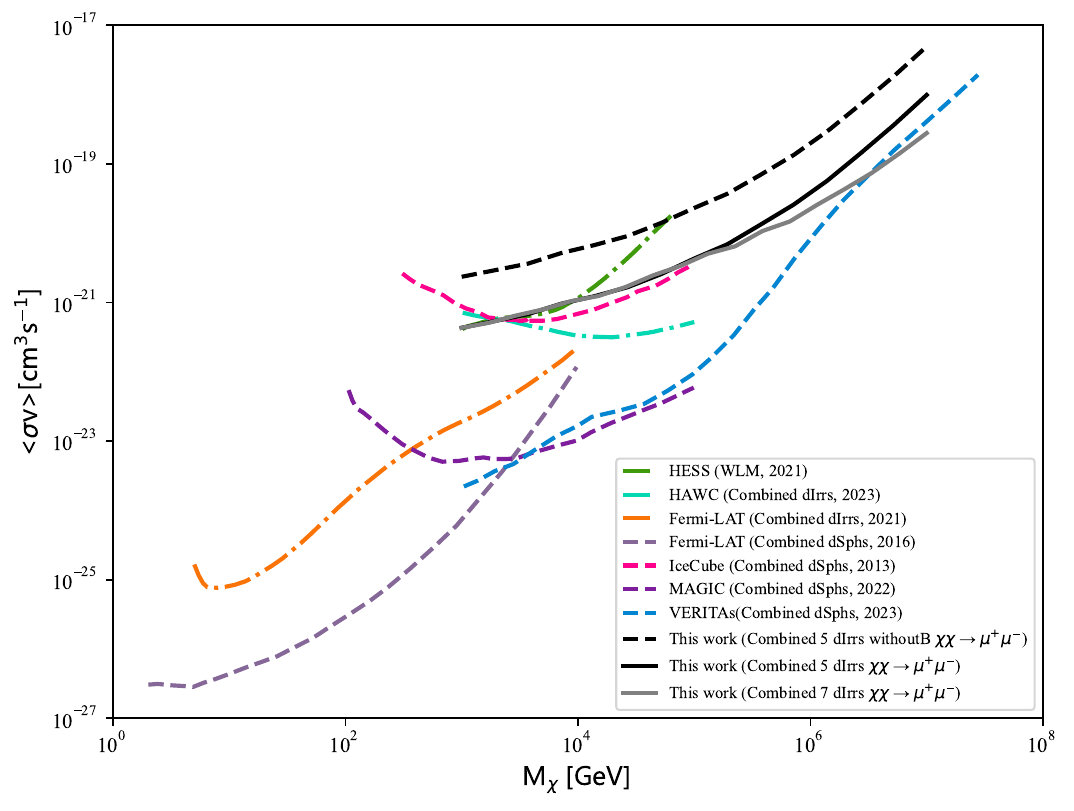}
    \caption{A comparison of our constraints (black solid and dashed lines) with other existing limits in the literature. The results plotted for comparison are based on the observations from IceCube (\citealt{2013PhRvD..88l2001A}), Fermi-LAT (\citealt{2015PhRvL.115w1301A,2021PhRvD.104h3026G}), HAWC (\citealt{2023ApJ...945...25A}), HESS (\citealt{2021PhRvD.103j2002A}), MAGIC (\citealt{2022arXiv220507055M}) and VERITAS (\citealt{2023ApJ...945..101A}).}
    \label{Figure:bijiao}
\end{figure}

Finally, in Fig.~\ref{Figure:bijiao} we compare the results of this analysis with the limits obtained in other works (\citealt{2013PhRvD..88l2001A, 2015PhRvL.115w1301A,2021PhRvD.104h3026G,2023ApJ...945...25A,2021PhRvD.103j2002A,2022arXiv220507055M,2023ApJ...945..101A}). 
The combined limits are stronger than the results from the HESS observations of WLM dIrr (\citealt{2021PhRvD.103j2002A}) at the DM masses above 10 TeV. Compared to other results given by analyzing dSphs, it can be seen that our constraints are weaker, mainly owing to the fact that dwarf irregular galaxies are more distant and thus have smaller J-factors. The advantage of our results, however, is that the J-factors of dIrr galaxies can be more accurately determined compared to dSphs and the constraints are thus more robust, as they are rotationally-supported galaxies. In contrast, as pressure-supported systems, dSphs have larger systematic uncertainties in the J-factor determination (\citealt{2013NewAR..57...52B}). However, for dIrrs, there exists another systematic uncertainty from the determination of the boost factor, as discussed in Sec.\ref{sect:2}. For this reason, here we also present the upper limits given by the joint analysis when the boost factor is not taken into account, and the results are shown as dashed black lines in Fig.~\ref{Figure:result} and Fig.~\ref{Figure:bijiao}.

 From the individual constraints in Fig.~\ref{Figure:result}, we can see that the strength of the constraints depends mostly on the values of the dIrr J-factors, so the uncertainty of our results comes primarily from the J-factor uncertainty, which we discuss below. The (statistical) uncertainty on the dIrr J-factors includes mainly the error on the DM profile parameters obtained in the fit to the RC, which depends on the quality of kinematical measurements. This error is $\sim$15\% for $\rho_0$ and $r_0$ and leads to a $\sim$75\% uncertainty in the astrophysical J-factor \citep{2018PhRvD..98h3008G}. In addition to this, there are some systematic uncertainties in the J-factor estimation, e.g. the choice of the model for the DM density profile causes a systematic uncertainty of $\sim$0.2 dex \citep{2021PhRvD.104h3026G}. The J-factors used in this paper, which are derived in Ref.~\citet{2018PhRvD..98h3008G}, assume a cored density distribution described by the Burkert profile ~\citep{1995ApJ...447L..25B}, and such a profile is supported by the high-resolution HI observations \citep{2011AJ....141..193O,2015AJ....149..180O} and can fit the rotation curves better \citep{2021PhRvD.104h3026G}. However, DM-only N-body cosmological simulations favor a cuspy Navarro-Frenk-White profile \citep{1997ApJ...490..493N}. 

The largest uncertainty in the predicted DM annihilation flux comes from the uncertain nature of the dark matter substructure (i.e., the uncertainty in the enhancement factor $B$). As can be seen from Fig.~\ref{Figure:bijiao}, whether the boost factor is considered or not, there is a factor of $\sim$5 difference in the results. In the case of considering the boost factor, different choices of the parameters of the subhalo population (e.g., the slope of the subhalo mass function, the minimal subhalo mass) lead to distinct (boosted) J-factors, as has been discussed in Sec.~\ref{sect:2} (see also Refs.~\citep{2021PhRvD.104h3026G,2017MNRAS.466.4974M,2017PhRvD..95f3531L}). Overall, for a dIrr of $10^{10}\,M_\odot$, the boost factor varies over a range of about 2.1 to 15.9 \citep{2017MNRAS.466.4974M}, and thus the constraints (with $B$) on the DM parameters have the same level of uncertainty.

\section{CONCLUSIONS}
Pressure-supported dSphs are often considered as one of the most promising targets for indirect DM searches because of their close proximity and negligible astrophysical background. The problem, however, is that the DM content of dSphs and their astrophysical J-factors are usually subject to significant systematic errors due to their kinematic uncertainty. In contrast, dIrrs are rotationally supported dwarf galaxies, which are also DM-dominated objects. Unlike dSphs, the kinematics of dIrr galaxies provide a good estimate of their DM halo density distribution, and hence of their astrophysical J-factors. Although there exists certain star formation activity within dIrrs, it has been shown that the astrophysical background component is negligible (\citealt{2018PhRvD..98h3008G}). They are therefore interesting noval targets for indirect detection of dark matter. In this work, we search for signals from UHDM annihilation in seven dIrrs using 10 years of IceCube neutrino data. In the calculation of the J-factors of the dIrrs, we use a Burkert profile to characterize their DM distribution and consider the boost factors due to the existence of subhalos. Also, considering the angular extension of dIrr galaxies in the sky, we analyze them as extended sources.

We do not detect any significant signal of DM annihilation in the directions of these 7 dIrrs. Based on the null results, we can place constraints on the model parameters of dark matter, since the emission from the annihilation should be detected if the annihilation cross section is large enough. Consequently, we derive the upper limits on the annihilation cross section at the 95\% confidence level for the DM with masses between 1 TeV and 10 PeV. For the single-source analysis, the dIrr DDO154 gives the strictest exclusion limits. We also compute the results of the joint analysis of the 5 dIrrs in the northern sky. The joint analysis gives stronger (by a factor of 2-3) constraints on the annihilation cross section compared to the single-source analysis.

In contrast to other experimental results, our constraints are equivalent to those given by the HAWC observations of dIrrs (\citealt{2023ApJ...945...25A}) and stronger than the constraints from the HESS observations of the WLM dIrr (\citealt{2021PhRvD.103j2002A}). We also find that the exclusion limits of dIrrs are weaker than but close to those of dSphs (especially for the latest VERITAS results (\citealt{2023ApJ...945..101A}) at PeV masses), considering the smaller J-factor uncertainties of dIrrs, our results therefore provide a valuable complement to the existing results.

\normalem
\begin{acknowledgements}
This work is supported by the National Key Research and Development Program of China (No. 2022YFF0503304), the National Natural
Science Foundation of China (No. 12133003) and the Guangxi Science Foundation (No. 2019AC20334).
\end{acknowledgements}
  
\bibliographystyle{raa}
\bibliography{bibtex}

\end{document}